Giant magnetic pumping of photovoltage and photocurrent using dielectric lossy material
Ye Wu, Amar Bhalla, Ruyan Guo
Department of Electric Engineering, University of Texas at San Antonio, One UTSA Circle, San Antonio, Texas 78249-0698, USA (Email: ruyan.guo@utsa.edu)

**Abstract**: The enhancement of photovoltage and photocurrent of material is fundamentally significant owing to the many interesting phenomena found and the potential applications. However, vast altering of magnitude-orders of photoelectricity has been technologically challenging. Here we report two dielectric materials $Li_2ZnSiO_4$ and $Li_2SiO_3$ showing high photovoltage and photocurrent tunability. When magnetic field increasing from 0.00015T up to 0.44T, it is found that 3850% of photovoltage tunability and 3841 % of photocurrent tunability in $Li_2ZnSiO_4$, and 132.8% of photovoltage tunability and 132.5% of photocurrent tunability in $Li_2SiO_3$. A simple model that considers the effects of spin mixing/interaction and magnetically-tunable charge gradient is used to explain this interaction between magnetic field and the photoelectricity. This result indicates a magnetic approach can be potentially used for energy efficiency.

Fossil fuels is certainly necessary for everyone's daily life, which could be used for car gasoline, heating, cooling or electricity generation. The fossil energy depletion has made the development of alternative energy and the corresponding energy harvest very important. Light energy is one type of the alternative energy that could be used to produce electrical power. Simply shining beams of light on certain material surface, researchers have been able to produce current and open circuit voltage. This effect is considered as photoelectricity, which is typically thought to be introduced by the generation of additional charge carriers on the crystal. [1] In general, the photocurrent is produced by the charge gradient ($j = \nabla n$). If the researchers are planning to increase the efficiency of the photocurrent generation, the first idea that comes to mind could be using pressure, temperature and magnetic field. Because the transport of charge carriers could be subjected to the external field interaction. One example would be using pressure to change the size of the material leading to resistance change. Another example would be increasing the temperature that could introduce heat on the atomic structure of a material and intensify the vibration of the atoms, cause more electron collisions and hence greater resistance to the current flow. The third example would be using magnetic field and hence introducing magnetic force to affect the movement of the charge carriers. However, it has been a great challenge to realize considerable change of photoelectricity. At the time of this article, several studies show a relatively low tunability of photocurrent through using the pressure, temperature and magnetic field. For example, it is reported that high pressure could shift the center of the photocurrent absorption spectrum[2] in titanium dioxide and photocurrent emission spectrum in $In_xGa_{1-x}N$ [3]. It is also reported that the photocurrent could be decreased 84% when temperature is decreased from 299K to 126K in polymer/fullerene bulk heterojunction solar cells[4]. In 2014, Z.G. Sheng et al. found 30% enhancement of photocurrent with the magnetic field up to 6T in manganite-based material[5]. They demonstrated that this effect is due to modulation of the bandgap in correlated electron oxides by the magnetic field.

While this strategies that using correlated electron systems to strengthen the photocurrent is effective given that their phase transitions are extremely sensitive to the external field and photocurrent is depended on the phase change, another effective strategy would be involving spin mixing and interaction. More generally, either electron spin or nuclear spin is subjected to the change of magnetic field, which could vastly change the electron spin transport of the whole material system and thus introducing spin polarization.

Given that the alkali atoms (e.g. Li atoms) based material system that includes various level structure, the lithium based composite is a promising candidate for study of magnetic field enhanced photocurrent and makes them an excellent platform in revealing the underlying material physics. Furthermore, some of them are typical dielectric materials which are more sensitive to optical field stimuli comparing to metals and semiconductors.[6]

Herein, we choose two lithium based dielectric material $Li_2ZnSiO_4$ and $Li_2SiO_3$ for our study of photocurrent and photovoltage under a magnetic field. We demonstrate that ultrahigh magnetic field enhancement of photovoltage and photocurrent could be achieved in these two dielectric lossy material, $Li_2ZnSiO_4$ and $Li_2SiO_3$. we report 3850% of photovoltage tunability with magnetic field 0.4T and 3841 % of photocurrent tunability with magnetic field 0.4T in $Li_2ZnSiO_4$, 132.8% of photovoltage tunability and 132.5% of photocurrent tunability in $Li_2SiO_3$. We propose a mechanism which considers the effects of electron/nuclear spin mixing and interaction and charge gradient under the magnetic field. This introduces large tunability and makes them suitable for practical application in optoelectronic technologies.

**Results.**

**Structure of the samples.**

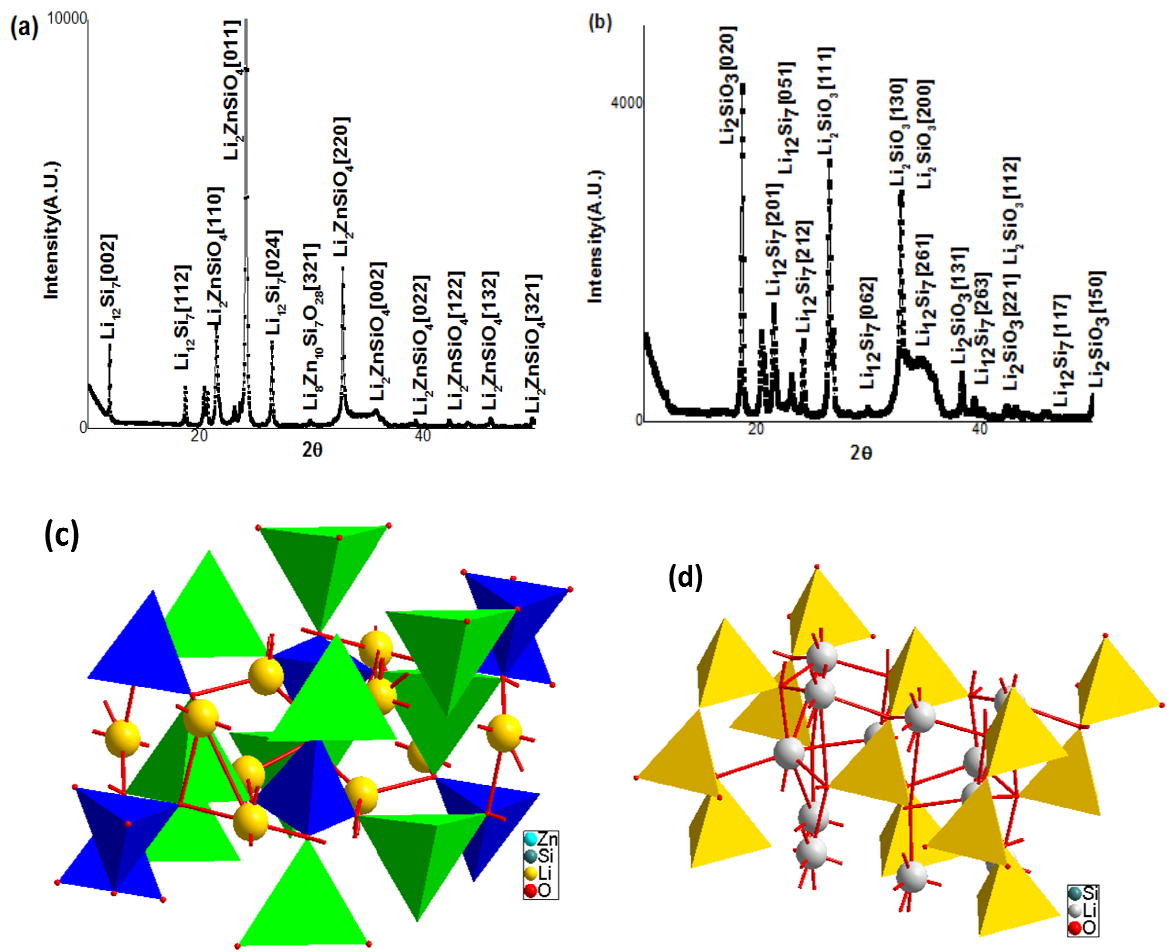

Figure.1 (a) XRD profile of $Li_2ZnSiO_4$ (b) XRD profile of $Li_2SiO_3$. (c) The unit cell of $Li_2ZnSiO_4$. (d)The unit cell of $Li_2SiO_3$.

The phases of the two samples were confirmed by XRD (Fig.1(a) and (b)). The lattice parameter of $Li_2ZnSiO_4$ is refined as, a=6.27Å, b=10.61Å, c=5.02Å, α=γ=90, β=90.45°. The space group is P 1 21/n 1.As shown in Fig.1(c), one $Zn^{+2}$ atom and four $O^{-2}$ atoms form a tetrahedron (green);one $Si^{+4}$ atom and four $O^{-2}$ atoms also form a tetrahedron (blue).The lattice parameter of $Li_2SiO_3$ is refined as a=9.47Å,b=5.39Å,c=4.63Å,α=β=γ=90°. The space group is C m c 21. As shown in Fig.1(d), one $Si^{+4}$ atom and four $O^{-2}$ atoms form a tetrahedron (yellow).It should be noted that several peaks in Fig.1(a) and Fig.1(b) were identified as the phases of $Li_{12}Si_7$, which should be due to strong reaction between the residue of the $Li^+$ with the silicon in the sample preparing process.

**Characterization of the materials.**

Admittance and dielectric spectroscopy were used to characterize the sample. Both samples show the capacitive behavior (Fig.2).The high loss tangent value (Fig.2(d)) indicates that both of them are lossy dielectric material.

The photo-voltage, -current, -resistance with respect to the optical power and time are shown in Fig.3. Here we used a 650nm laser to shine the samples and did not apply any magnetic field. Several characteristics are summarized as followed.

The photovoltage and photocurrent are increased with the increasing of the optical power(Fig.3 (a) and (b)). The photoresistance shows small fluctuation (<1%)(Fig.3(c)).When the optical power is increased (400mW), the photocurrent, photovoltage, photoresistance show small deviation (<10%) with time (Fig.3 (d)-(e)).

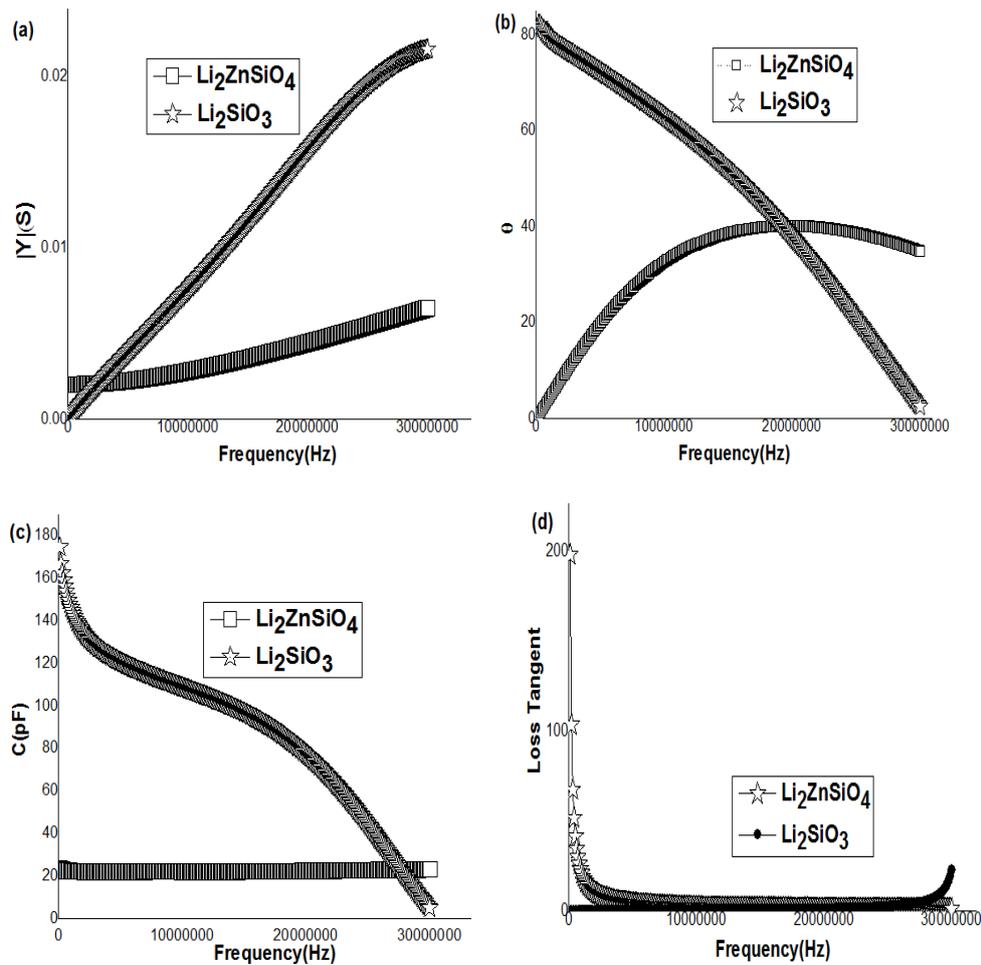

Figure 2 (a) admittance ,(b) phase (c) capacitance and (d) loss tangent vs. frequency.

**Photoelectricity enhancement by magnetic field.**

The photovoltage and photocurrent of $Li_2ZnSiO_4$ and $Li_2SiO_3$ were found to be dependent on the magnetic field (Fig.4). It is found that 3850% of photovoltage tunability with magnetic field 0.44T and 3841 % of photocurrent tunability with magnetic field 0.44T in $Li_2ZnSiO_4$ composite material,132.8% of photovoltage tunability and 132.5% of photocurrent tunability with magnetic field 0.44T in $Li_2SiO_3$ compounds. Here photovoltage and photocurrent tunability is defined as

$$\frac{V(B>0.00015T)-V(B=0.00015T)}{V(B=0.00015T)} \text{ and } \frac{I(B>0.00015T)-I(B=0.00015T)}{I(B=0.00015T)} \text{ respectively,}$$

where V(B>0.00015T) is the photovoltage when magnetic field B>0.00015T , V(B=0.00015T) is the photovoltage when magnetic field B=0.00015T , I(B>0.00015T) is the photocurrent when magnetic field B>0.00015T , I(B=0.00015T) is the photocurrent when magnetic field B=0.00015T.

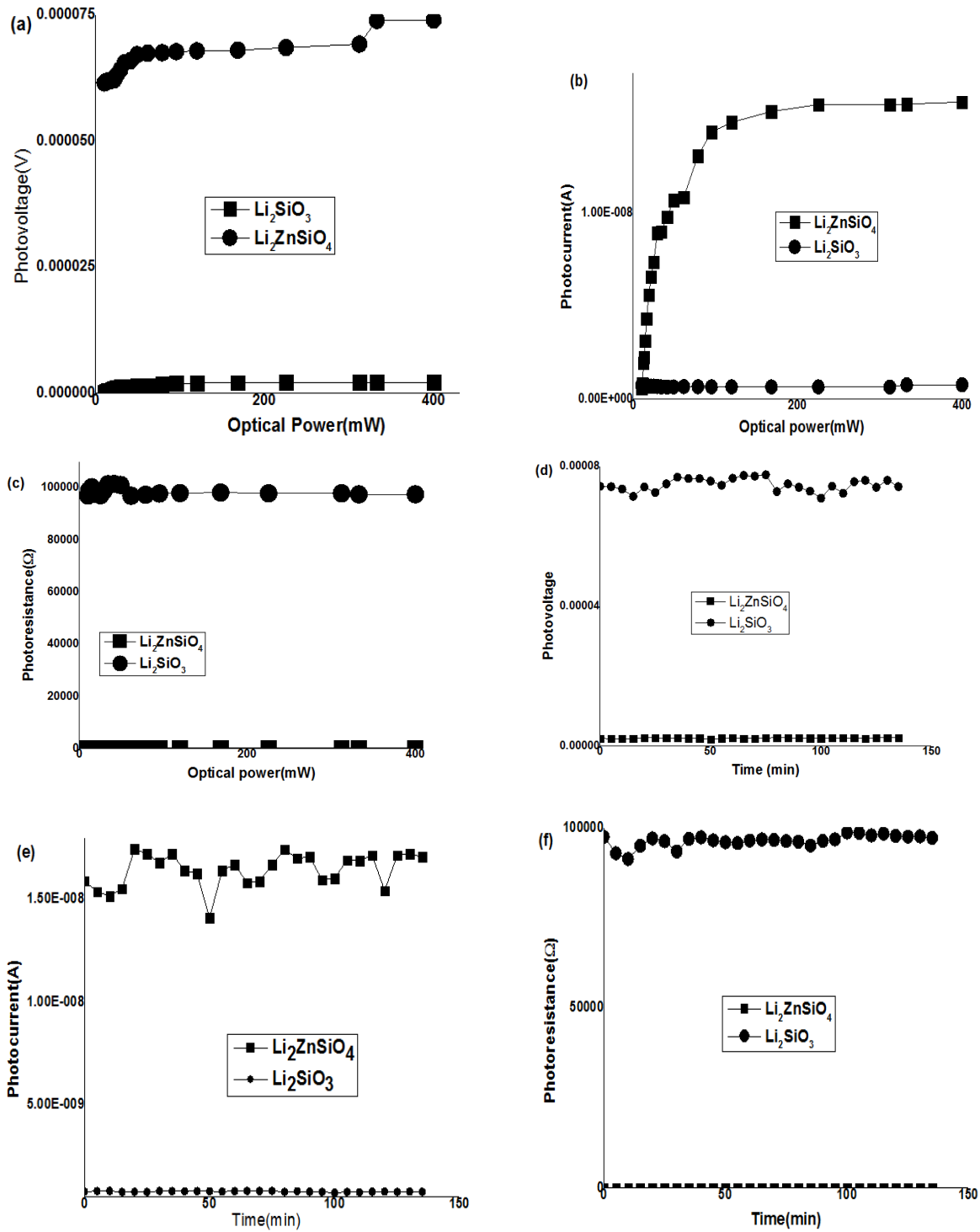

Figure 3 Photovoltage, photocurrent and photoresistance dependence on optical power and time.

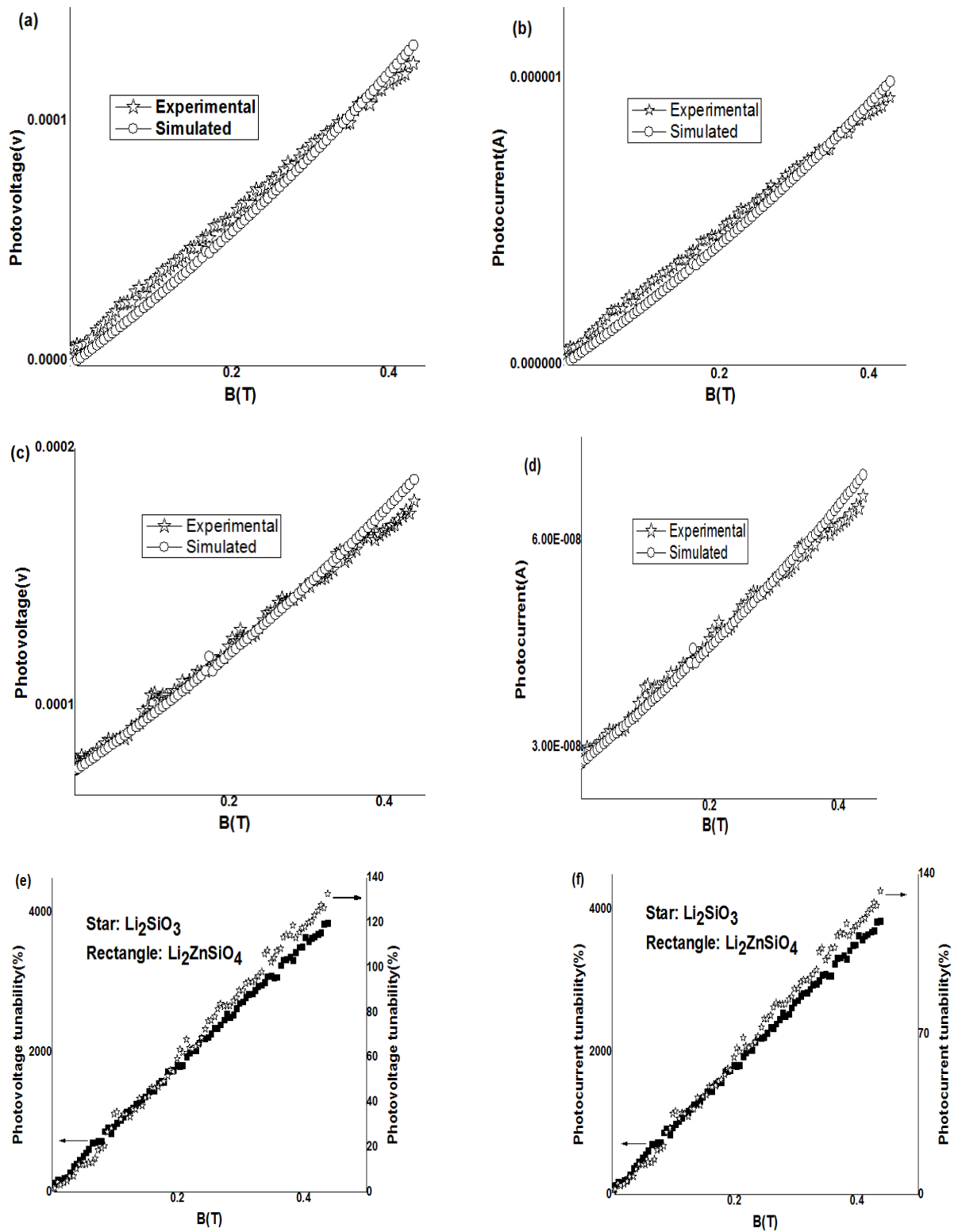

Figure 4. (a) photovoltage and (b) photocurrent of $Li_2ZnSiO_4$. (c) Photovoltage and (d) photocurrent of $Li_2SiO_3$. (e) Photovoltage and (f) photocurrent tunability of $Li_2ZnSiO_4$ and $Li_2SiO_3$.

**Theoretical model.**

For the following, it would be important to discuss the mechanism of magnetic tunability of photoelectricity in these two dielectric materials. These two materials contain the element of lithium,oxygen,silicon and zinc. It should be noted that the element of lithium is paramagnetic, oxygen paramagnetic,silicon diagmagnetic and zinc diagmagnetic. Therefore what is open to the discussion is about the reason why the seemingly nonmagnetic elements of Li,O,Si and Zn form materials showing magnetic feature. We assume a scenario that is similar to what have been found in magnetic field effect of organic molecules, where nonmagnetic organic molecules showing magneto-photoluminescence, absorption and conductance due to the spin mixing of hydrogen.[7][8][9][10][11] We assume a process of electron and nuclear spin mixing and interaction existing in these two materials. This process could be originally generated from Li, whose electric structure is similar to hydrogen. The generation of photocurrent within the magnetic field could be understood as the effect of magneto-opto-electric coupling. We consider that the electron is excited to different spin states due to laser excitation, while at the same time, spin polarization is constructed with different spin states of either up-spin or down-spin. The current is thus produced by the spin polarization. From the quantum mechanical point of view, the Hamitonian of the system associated with this process should contain the basic energy, the energy related to the laser-material interaction and the hyperfine structure, which would be discussed in detail in the following.

We firstly look into the electron /nuclear spin mixing and interaction effect for the photo-charge transport in the magnetic field. More specifically, we introduce the parameter of spin polarization α, which contributes to the photocurrent in the form of [12]

$$j_1 = erl(sp - \frac{s - \alpha n}{\tau}). \qquad (1)$$

Here $r$ is the recombination rate, $p$ is the hole concentration, $s$ is the spin density, $\alpha$ is the spin polarization, $\tau$ is the spin relaxation time.

With the basic form of the photocurrent given, we now turn to derivation of the spin polarization. As mentioned above, we consider the total Hamiltonian is the sum of basic Hamiltonian $H_b$, the magnetic field interaction with the hyperfine structure $H_{hf}$, and the material-laser interaction $H_{laser}$. Thus this Hamitonian has reflected the process of magneto-opto-electric coupling.

The basic Hamiltonian is given by $H_b = \frac{p_0^2}{2m} + \hbar\omega_0(|e_1\rangle\langle e_1| + |e_2\rangle\langle e_2| + |e_3\rangle\langle e_3| + |e_4\rangle\langle e_4|)$, (2)

which is the sum of the kinetic energy and the internal energy. Here excited levels are defined as,

$|e_1\rangle = |s = \frac{1}{2}, I = \frac{1}{2}\rangle$ ,(3)  $|e_2\rangle = |s = \frac{1}{2}, I = -\frac{1}{2}\rangle$ ,(4)  $|e_3\rangle = |s = -\frac{1}{2}, I = \frac{1}{2}\rangle$ , (5)  and

$|e_4\rangle = |s = -\frac{1}{2}, I = -\frac{1}{2}\rangle$ .(6)

The magnetic and hyperfine interaction is given by

$$H_{hf} = g_e\mu_B BS_Z - g_N\mu_N BI_z + hA\vec{I}\vec{S}. \qquad (7)$$

Here the magnetic field is supposed to be along the z-direction. $g_e$ is electron g-factor, $\mu_B$ is the Bohr magneton, $g_N$ is nuclear g-factor, $\mu_N$ is the nuclear magneton, $S_z$, S are the electron spin, I is the nuclear spin, A is the hyperfine tensor.

The energy from the laser-material interaction is given by $H_{laser} = -e\vec{r}\vec{E}$, (8)

where $\vec{r}$ is the electric dipole operator, $\vec{E}$ is Electric component of the optical wave. From Eq.(2),(7) and (8), we can obtain every element of the total Hamiltonian matrix and therefore calculate the eigenvalues of the total Hamiltonian through the roots of the characteristic equation as

$$\lambda_1 = \frac{p_0^2}{2m} + \hbar\omega_0 + \frac{g_e\mu_B B}{2} - \frac{g_N\mu_N B}{2} + \frac{hA}{4}, (9) \quad \lambda_2 = \frac{p_0^2}{2m} + \hbar\omega_0 + \frac{g_e\mu_B B}{2} + \frac{g_N\mu_N B}{2} - \frac{hA}{4}, (10)$$

and $\lambda_{3,4} = \frac{-g_e\mu_B B}{2} + \frac{p_0^2}{2m} + \hbar\omega_0 \pm \sqrt{\Delta}$, (11),

where
$$\Delta = (\frac{p_0^2}{m} + 2\omega_0\hbar - g_e\mu_B B)^2 - 4\left(\frac{p_0^2}{2m} + \omega_0\hbar - \frac{g_e\mu_B B}{2} - \frac{g_N\mu_N B}{2} - \frac{hA}{4}\right)$$
$$\left(\frac{p_0^2}{2m} + \omega_0\hbar - \frac{g_e\mu_B B}{2} + \frac{g_N\mu_N B}{2} + \frac{hA}{4}\right) + 4\hbar^2\omega_{34}\omega_{43}$$
(12)

Here $\omega_{ij} = \frac{\vec{d}_{ij}\vec{E}}{\hbar}$ (13) and $\vec{d}_{ij}$ is the transition dipole moment for i-j transition.

The eigenstates function can be written as $|e\rangle = \lambda_1|e_1\rangle + \lambda_2|e_2\rangle + \lambda_3|e_3\rangle + \lambda_4|e_4\rangle$ (14)

We suppose $|e_1\rangle$, $|e_2\rangle$, $|e_3\rangle$ and $|e_4\rangle$ are 4 spin states with the number of atoms given by $N_1$, $N_2$, $N_3$ and $N_4$, respectively. Using Boltzmann factor, we can express $N_2 = N_1\exp\frac{\lambda_1-\lambda_2}{kT_2}$, (15)

$N_3 = N_1\exp\frac{\lambda_1-\lambda_3}{kT_3}$ (16) and $N_4 = N_1\exp\frac{\lambda_1-\lambda_4}{kT_4}$ (17), where T2, T3 and T4 are thermodynamic temperature, k is Boltzmann's constant.

The electron spin polarization is expressed as $\alpha = \frac{(N_1+N_2)-(N_3+N_4)}{N_1+N_2+N_3+N_4}$. (18)

Plugging Eq.(15)-(17) into Eq.(18), we can derive

$$\alpha = \frac{1+\exp\frac{\lambda_1-\lambda_2}{kT_2} - \exp\frac{\lambda_1-\lambda_3}{kT_3} - \exp\frac{\lambda_1-\lambda_4}{kT_4}}{1+\exp\frac{\lambda_1-\lambda_2}{kT_2} + \exp\frac{\lambda_1-\lambda_3}{kT_3} + \exp\frac{\lambda_1-\lambda_4}{kT_4}}$$
(19)

For simplicity, we assume $\Delta \approx (\frac{p_0^2}{m} + 2\omega_0\hbar - g_e\mu_B B)^2$ (20) and $kT_1 \approx kT_2 \approx kT_3 \approx kT_4 \approx kT$ (21).

Using Taylor series expansion for exponential function, we derive a simple form of $\alpha$ as

$$\alpha = \frac{B - \dfrac{hA}{4g_e\mu_B}}{B(\dfrac{g_N\mu_N}{g_e\mu_B} - 1) - \dfrac{3hA}{4g_e\mu_B} - \dfrac{4kT}{2g_e\mu_B}} \tag{22}$$

Plugging Eq.(22) into Eq.(1), the carriers current is derived as

$$j_1 = erlsp - \frac{erls}{\tau} + \frac{er\ln}{\tau}\frac{B - \dfrac{hA}{4g_e\mu_B}}{B(\dfrac{g_N\mu_N}{g_e\mu_B} - 1) - \dfrac{3hA}{4g_e\mu_B} - \dfrac{4kT}{2g_e\mu_B}} = C_0 + \frac{BC_1 - C_2}{BC_3 - C_4} \tag{23}$$

where $C_0, C_1, C_2, C_3$ and $C_4$ are the constants.

While we have microscopically derived photocurrent dependence on the magnetic field using the interaction between the magnetic field and electron/nuclear spin, we will continue to discuss the photocurrent in a classical way.

The diffusion current has the form of $j_2 = eD_n \dfrac{\partial n}{\partial l}$, (24) where $D_n$ is the diffusion coefficient, $e$ is the electron charge, $l$ is the spatial length, n is the carrier density

It suggests that if we can derive the dependence of charge gradient on the magnetic field and optical wave, we could establish the relationship between the photocurrent and the magnetic field.

Indeed, this effect could be considered as carrier concentration manipulation by magnetic injection or "pumping". If we can alter the carrier concentration by magnetic transport and injection somewhere on the crystal surface, there will be a diffusion of the excess carriers throughout the crystal.

We conjecture that the charge will experience the magnetic force, repulsive force, attractive force. Using a semiclassical derivation [13], the response of the charge carriers to the magnetic field interaction could be described by the force balance equation as

$$\frac{\hbar^2}{m}\pi^{4/3}3^{-1/3}n^{-1/3}\frac{dn}{dl} = e(E_0 + ke\int_0^l n(\gamma)d\gamma) + eBv \tag{25}$$

Here, m is the electron mass, $\hbar$ is the plank constant, n is the charge density, e is element charge, $E_0$ is the electric field component of the optical wave. $l$ is the sample length. $v$ is the charge moving velocity and k is the constant that depends on the material permittivity.

Eq.(25) has established the relationship between the charge gradient and the magnetic field / optical wave.

Again using a similar approach in Ref.[13], we can generate the charge gradient dependence on the magnetic field.

$$\frac{\partial n}{\partial l} = \frac{-6B\sqrt{C_5 n^{5/3} - C_6}\sqrt{C_7 n^{1/3} - C_8}}{-4Bn^{-1/3}\sqrt{C_7 n^{1/3} - C_8} + 9n^{-5/3}\sqrt{C_5 n^{5/3} - C_6}} \qquad (26)$$

Where $C_5, C_6, C_7, C_8$ are the constants defined in ref.[13]

The diffusion current is calculated as,

$$j_2 = eD_n \frac{\partial n}{\partial l} = \frac{-6BeD_n\sqrt{C_5 n^{5/3} - C_6}\sqrt{C_7 n^{1/3} - C_8}}{-4Bn^{-1/3}\sqrt{C_7 n^{1/3} - C_8} + 9n^{-5/3}\sqrt{C_5 n^{5/3} - C_6}}, \qquad (27)$$

where $D_n$ is the diffusion coefficient, $e$ is the electron charge, $l$ is the spatial length, n is the carrier density. Finally, we consider the total photocurrent is the combination of $j_1$ and $j_2$, which is $j=k_1 j_1 + k_2 j_2$. (28). Here $k_1$ and $k_2$ are the coupling factor for $j_1$ and $j_2$, respectively.

Plugging Eq.(23) and Eq.(27) into Eq.(28) and reorganizing it, the total photocurrent is found to be

$$j = k_1 C_0 + k_1 \frac{BC_1 - C_2}{BC_3 - C_4} + k_2 \frac{6BeD_n\sqrt{C_5 n^{5/3} - C_6}\sqrt{C_7 n^{1/3} - C_8}}{4Bn^{-1/3}\sqrt{C_7 n^{1/3} - C_8} - 9n^{-5/3}\sqrt{C_5 n^{5/3} - C_6}} \qquad (29)$$

If some parameters in Eq.(29) are adsorbed into the constants, we can derive a simple form of Eq.(29) as

$$j = C_0 + \frac{BC_1 - C_2}{BC_3 - C_4} + \frac{BC_5}{BC_7 - C_6} \qquad (30)$$

Based on Eq.(30) and assuming the resistance value is experimental determined, we can simulate the photovoltage and photocurrent under magnetic field. Fig.3 (a)-(d), compares the experimental and simulated data of the photovoltage and photocurrent of $Li_2ZnSiO_4$ and $Li_2SiO_3$ with the magnetic field applied. Although a very precise quantitative analysis of our data in terms of theoretical calculations capturing all the details of the composite material investigated is beyond the present scope, we do show the trend of the photoelectricity dependence on the external magnetic field. This simple model simulates the important characteristics of the highly magnetically tunable photovoltage and photocurrent in these two materials.

**Discussion**

More generally, we see magnetic field as a new and efficient tool for dynamical driving of photoelectricity. The photovoltage and photocurrent of $Li_2ZnSiO_4$ and $Li_2SiO_3$ both increase with the magnetic field increasing. When B=0.44T, we find that 3850% of photovoltage tunability and 3841 % of photocurrent tunability in $Li_2ZnSiO_4$, 132.8% of photovoltage tunability and 132.5% of photocurrent tunability in $Li_2SiO_3$. In order to explain the interaction of photoelectricity tunability with magnetic field, we present a mechanism which takes account of the effects of spin polarization and magnetically-tunable charge gradient and on charge transport. This adds to the new knowledge of the magneto-opto-electric coupling in material. The ultrahigh photoelectricity tunability with low magnetic field of these two materials shows a possible route to introduce novel tunable photoelectric devices.

**Methods**

**Sample Preparation.**

The samples used in the study were produced by polymer assisted deposition.[14]

The materials we used were lithium nitrate(Fisher Scientific co.), zinc oxide (J.T.Baker chemical Co.),hydrochloric acid (Duda Diesel LLC) ,and polyethyleneimine (Alfa Aesar).

For the deposition of $Li_2SiO_3$ film, $LiNO_3$ (0.24g) was dissolved in HCl(32%,0.28g).Then polyethyleneimine liquid (1.05g) was added into the Li stock solution. The Li precursor solution was spin-coated on a silicon substrate. The coated film was sintered at 900 ℃ for 7 hours.

This synthesized $Li_2SiO_3$ film could be further used for the deposition of $Li_2ZnSiO_4$ material. The ZnO (0.21g) was dissolved in HCl (32%,0.63g). Then polyethyleneimine liquid(1.03g) was added into the Zn stock solution. The Zn precursor solution was spin-coated on the $Li_2SiO_3$ film. The coated film was sintered again at 1100 ℃ in air for 10 hours.

**Sample crystal structure refinement and characterization.**

The sample structure was refined and drawn by GSAS software package. [15] Then the refined unit cell is drawn by DRAWxtl.[16] (Figure 1 (c) and (d)). The magnetic field was provided by a lab-built DC magnet. In measurements of both samples, the magnetic field was applied parallel to the plane of the substrate. The photocurrent and photovoltage were produced by irradiating the samples with a photodiode (center wavelength: 650nm, output power: 400mW).The electrical characterization was carried by a multimeter(Fluke 8846A). The dielectric characterization was done by using an impedance analyser (Hp 4194A). Ellipsometer(Gaerther L116B) was used to determine the thickness of the film. The thickness of $Li_2SiO_3$ film is 20μm; the thickness of $Li_2ZnSiO_4$ film is 52μm.

ACKNOWLEDGMENTS

This work was supported by NSF and INAMM.The authors thank Dr.Hailong Wang and Dr. Wei Yang in UTSA Chemistry for many fruitful discussions.